\documentclass[lettersize,journal]{IEEEtran}
\usepackage{amsmath,amsfonts}
\usepackage{algpseudocode}
\usepackage{algorithm}
\usepackage{blindtext}
\usepackage{array}
\usepackage[caption=false,font=normalsize,labelfont=sf,textfont=sf]{subfig}
\usepackage{textcomp}
\usepackage{stfloats}
\usepackage{url}
\usepackage{verbatim}
\usepackage{graphicx}
\usepackage{diagbox}
\begin{document}
\title{Link Prediction and Unlink Prediction on Dynamic Networks}

\author{Christina Muro\IEEEauthorrefmark{1}, Boyu Li\IEEEcompsocitemizethanks{\IEEEcompsocthanksitem\IEEEauthorrefmark{1}The first two authors contribute equally.}, Kun He,\IEEEauthorrefmark{2}\IEEEcompsocitemizethanks{\IEEEcompsocthanksitem\IEEEauthorrefmark{2}Corresponding author.}
\thanks{C. Muro, B. Li and K. He are with the School of Computer Science and Technology, Huazhong University of Science and Technology, Wuhan, China (E-mail: \{christinamuro, afterslby, brooklet60\}@hust.edu.cn)}}


\markboth{IEEE Transactions,~Vol.~??, No.~?, ????~2022 }
{Shell \MakeLowercase{\textit{et al.}}: A Sample Article Using IEEEtran.cls for IEEE Journals}


\maketitle

\begin{abstract}
Link prediction on dynamic networks has been extensively studied and widely applied in various applications. However, temporal unlink prediction, which also plays an important role in the evolution of social networks, has not been paid much attention. Accurately predicting the links and unlinks on the future network greatly contributes to the network analysis that uncovers more latent relations between nodes. In this work, we assume that there are two kinds of relations between nodes, namely long-term relation and short-term relation, and we propose an effective algorithm called LULS for temporal link prediction and unlink prediction based on such relations. Specifically, for each snapshot of a dynamic network, LULS first collects higher-order structure as two topological matrices by applying short random walks. Then, LULS initializes and optimizes a global matrix and a sequence of temporary matrices for all the snapshots by using non-negative matrix factorization (NMF) based on the topological matrices, where the global matrix denotes long-term relation and the temporary matrices represent short-term relations of snapshots. Finally, LULS calculates the similarity matrix of the future snapshot and predicts the links and unlinks for the future network. Additionally, we further improve the prediction results by using graph regularization constraints to enhance the global matrix, resulting that the global matrix contains a wealth of topological information and temporal information. The conducted experiments on real-world networks illustrate that LULS outperforms other baselines for both link prediction and unlink prediction tasks.
\end{abstract}


\section{Introduction}
\IEEEPARstart{L}{ink} prediction is one of the fundamental problems that predicts whether two disconnected nodes in a network are likely to have a link \cite{kumar2020link}. It is useful in a wide variety of applications, such as recommendation \cite{mat_fac_tec}, network reconstruction \cite{a_rev_of_rel}, and protein-protein interaction \cite{net_bas_pre_of}. These networks always have dynamic nature, indicating that nodes and links can be added or removed with the evolution of networks \cite{evo_net_ana}. It is inspiring and, in some respects, difficult to study networks at the level of individual edge formation or removal, especially in the dynamic scenario. Consequently, it still needs to be further studied on understanding the mechanisms by which such networks evolve at the level of individual edges.

Link prediction has been studied extensively in recent years~\cite{kumar2020link, a_sur_of_link, link_pre_in_soc}, and there are two primary independent scenarios for predicting unknown links in the networks, forecasting of missing relationships and projection of future relationships. Although real-world networks have a highly dynamic structure, most of the previous link prediction studies focus on static networks \cite{the_link_pre}. In such scenario, it is anticipated  that once two nodes are tied, they mostly remain tied (e.g., in the Facebook network, once two people are connected, they rarely break up).  As a result, the link prediction problem focuses on identifying new links. On the other hand, for many networks, additional temporal information, such as the time of link creation and deletion as well as node addition and deletion, is available over a time interval, which is beneficial to understand the structure of the network. Inferring links in dynamic networks is more challenging because dynamic networks are often subjected to short-period changes due to noise. Furthermore, due to changes in the overall environment, some dynamic networks may endure long-term shifts \cite{chen2018exploiting}.

Existing dynamic link prediction methods consider only either the structure of the network~\cite{the_link_pre, deng2016latent, dunlavy2011temporal} or the temporal information~\cite{rossi2013modeling, sarkar2005dynamic}. However, these families of approaches have several limitations. First, real-world networks are generally sparse and have partially observable links, resulting that the approaches based on only structure information may perform poorly. Second, the methods considering temporal information alone may lack insight provided by the other model. It is thus essential to use both topological and temporal features to comprehend complex behaviours of the dynamic network. Up to now, only a few studies~\cite{chen2018exploiting, yu2017temporally} have utilized both types of information together for the dynamic link prediction task. However, these methods fail to consider the intrinsic geometric structure of the data~\cite{cai2008non} resulting in lacking discriminative information for prediction.

Additionally, unlink prediction, which attempts to predict whether a previously occurred relationship will disappear in the future, is an important fundamental problem related to link prediction. Several studies~\cite{fra_onl_rel, loo_fri_on_fac, the_imp_of_net} reveal that the probability of relationship to persist or to-be formed increases if a node pair has a high number of common neighbors as well as high transitivity through third parties. Furthermore, they also divulge that the probability of an edge to befall occurs between edges with low resemblance, fewer communal neighbors, and typically of low transitivity. These findings illustrate that the processes guiding link creations are negatively correlated with those guiding link removals, which signifies the importance of unlink prediction tasks. However, up until now, the link prediction problem has been heavily studied~\cite{the_link_pre, pre_and_ide_mis, pre_mis_lin}, while its counterpart, unlink prediction problem, although reporting a high proportion of link changes~\cite{the_bur_dyn_of}, is rarely studied~\cite{sem_unl_pre}.

In this work, we assume that the relations in the dynamic networks can be divided into two categories, long-term relation and short-term relation. The long-term relation represents a certain kind of stable relation, while the short-term relation denotes a sort of temporary relation. For instance, in the collaboration network, the authors in the same lab maintain a stable relation, and they may always collaborate on publishing papers. While the authors in the different labs or even focusing on different research fields have a temporary relation such that they only cooperate occasionally. As a result, these two kinds of relations determine the states of links and unlinks in the future network to a large extent. 

Based on such assumption, we propose a new algorithm for the \textbf{l}ink prediction and \textbf{u}nlink prediction with \textbf{l}ong-term relation and \textbf{s}hort-term relation, termed LULS. LULS utilizes the method of non-negative matrix factorization (NMF) that embeds nodes into two kinds of representation matrices containing both structural similarity and temporal information. More specifically, for each snapshot of network, LULS collects higher-order topological information as matrices by performing two random walk methods, which are light lazy random walk~\cite{Krylov_sub_app} and a new variant of random walk called modified light lazy random walk. Then, LULS initializes and optimizes a global matrix and a temporary matrix for each snapshot by using the method of NMF, where the global matrix represents the long-term relation that stores the temporal information and is shared by all the snapshots, while the series of temporary matrices denotes the short-term relation that only stores the topological information within the corresponding snapshot. Additionally, LULS also incorporates the geometric structure by using graph regularization constraints to enhance the global matrix. Therefore, the global matrix contains a wealth of both topological information and temporal information, which can further improve the prediction results. Afterwards, LULS calculates the similarity matrix of the future snapshot for the temporal predictions of links and unlinks based on the global matrix and temporary matrices. The experiments conducted on real-world networks illustrate that LULS outperforms other baselines on AUC for both tasks of link prediction and unlink prediction.

\section{Related Work}
Link prediction in static network has been extensively studied~\cite{kumar2020link}. On the other hand, for many real-world networks, additional temporal information is available overtime interval, and the network build from such data can be represent as a dynamic network. One feasible strategy to model the dynamic network is to explore the network topological properties, which indicates that any two nodes close to each other in the network are likely to form a link in the near future. For example, Rahman et al.~\cite{link_pre_in_dyn} capture the topology of dynamic networks by graphlets, where graphlet transitions between different timestamps are coded in a feature vector and can be used by supervised learning. Gunes et al.~\cite{link_pre_usi} apply the ARIMA model on the series of node similarity scores to predict links in the next period based on the previous time series data. Moradabadi et al.~\cite{moradabadi2017novel} predict the future connections between pairs of nodes by using the previous snapshot as input. Besides, many approaches based on the matrix factorization approach has also been proposed~\cite{acar2009link, deng2016latent, dunlavy2011temporal, yu2017temporally, zhu2016scalable}. Matrix factorization has the advantages of generating embedding which are simple to interpret. The main idea is to learn a latent low-dimensional vector representation for each node and the nodes close to each other in the low-rank space are similar. However, most of the real-world networks are sparse, so that the methods considering only the structural characteristics of the network may perform poorly.

Moreover, there are some approaches that utilizes temporal information, which reveals the relationship between the current snapshot and the previous snapshots, to predict links on dynamic networks~\cite{chi2007evolutionary, deng2016latent, lin2008facetnet, rossi2013modeling, yu2017temporally, yu2017link}. The main idea is the temporal smoothness, which assumes that the embeddings from the current snapshot should not deviate dramatically from the previous snapshots \cite{chi2007evolutionary}. Although the approaches utilizing structural or temporal information alone have shown an encouraging performance, the combination of topological and temporal information always provides further insights that are missed with the use of single modeling. LIST~\cite{yu2017link} is proposed which exploits structural and temporal information and characterizes network information using time function to predict links on evolving networks. STEP~\cite{chen2018exploiting} is presented to exploit structural and temporal information for prediction and characterize network evolution using a global transition matrix to reflect different types of evolutionary patterns. Despite their good predictive performance, these approaches fail to consider the intrinsic geometric structure of the data, lacking discriminative information for prediction. Other approaches such as those based on graph neural network (GNN)~\cite{chen2019lstm, yang2019advanced,lei2019gcn} requires the node features, which may not always be available or may produce embedding with limited interpretability.

In additional to link prediction, unlink prediction also plays an important role in network evolution. For example, although some temporary relationships among people in an online social network are formed during a short-term, some of these relationships are likely to decay or even disappear in the future. Until now, there are only few studies address the problem. Preusse et al.~\cite{str_dyn_of_kno} use various network structural features extracted from a knowledge network to predict the disappearance of links. Oliveira et al.~\cite{sem_unl_pre} propose the method combining both the topology information and the information of individuals (semantic metrics) on evolving networks to predict the disappearance of links.

Comparing to the previous link prediction and unlink prediction methods, LULS has several advantages. First, to preserve the higher-order topological information as much as possible, LULS applies two random walk methods on each snapshot, so that after optimization, the global matrix and the temporary matrices include more and better topological information. Second, LULS utilizes NMF to optimize a global matrix and a series of temporary matrices to reconstruct the matrices obtained by random walk methods. On the one hand, the global matrix is optimized by all the snapshots, resulting that the global matrix preserves a wealth of temporal information (i.e., long-term relation). On the other hand, each temporary matrix contains the specific topological information of the corresponding snapshot that accurately preserves the short-term relations between nodes. Therefore, the topological information of future snapshot can be precisely reconstructed. Last but not least, LULS also optimizes the global matrix by using graph regularization constraints which makes the global matrix preserve more topological information that further improves the prediction results.

\section{Preliminaries}
In this section, we first give the problem definition, and then briefly introduce the random walk method. Important symbols used in this paper are listed in Table \ref{tab_sym}.

\begin{table}[!t]
	\caption{Symbols and definitions.}
	\label{tab_sym}
	\centering
	\begin{tabular}{|c|c|}
		\hline
		\textbf{Symbol} & \textbf{Definition} \\
		\hline
		$G_t = (V_t, E_t)$ & the snapshot $G_t$ with node set $V_t$ and edge set $E_t$ \\
		\hline
		$N$ & the final timestamp \\
		\hline
		$\textbf{W}_t$, $\textbf{H}_t$ & the similarity matrices for $G_t$ \\
		\hline
		$\textbf{U}$ & the global representation matrix \\
		\hline
		$\textbf{V}_t$ & the temporary representation matrix for $G_t$ \\
		\hline
		$m$ & the dimension of node representation \\
		\hline
		$\theta$ & decay weight\\
		\hline
		$\lambda$ & smoothness weight\\
		\hline
		$\gamma$ & control the importance of constraint terms \\
		\hline
		$k$ & the steps of random walk \\
		\hline
	\end{tabular}
\end{table}

\subsection{Problem Definition}
Consider an undirected and unweighted graph $G=(V,E)$, where $V$ and $E$ denote the set of observed nodes and edges, respectively. Let $\textbf{A} \in [0,1]^{|V| \times |V|}$ be the associated adjacency matrix, $\textbf{D}$ be a diagonal matrix of node degrees, and $\textbf{I}$ represents the identity matrix. A dynamic network is represented as a sequence of snapshots of graph $\mathcal{G}=\{G_1, G_2, \cdots, G_N\}$, where $N$ is the final timestamp. We denote $G_t=(V_t,E_t)$ as the graph at timestamp $t$ ($1 \le t \le N$). For simplicity, we assume the set of nodes do not change over snapshots, i.e., $V_{t_i}=V_{t_j}$, for any $t_i, t_j \in \{1, \cdots, N\}$, indicating that we ignore newly added and removed nodes. However, edges do appear and disappear in snapshots over the timestamps. Our goal is to predict the edges that will be added or removed in $G_{N+1}$ based on the previous observed snapshots in $\mathcal{G}$.

\newtheorem{definition}{Definition}
\begin{definition}[Dynamic Link Prediction]
	Given a sequence of snapshots of graph $\mathcal{G}=\{G_1, G_2, \cdots, G_N\}$, where $N$ is the final timestamp, all the snapshots are with the same set of nodes but some edges may emerge or disappear along the sequence. For any pair of unconnected nodes $u$ and $v$ in $G_{N}$, the link prediction task on dynamic network aims to predict whether $u$ and $v$ will have a link in $G_{N+1}$.
\end{definition}

\begin{definition} [Dynamic Unlink Prediction]
	Given a sequence of snapshots of graph $\mathcal{G}=\{G_1, G_2, \cdots, G_N\}$, where $N$ is the final timestamp, all the snapshots are with the same set of nodes but some edges may emerge or disappear along the sequence. For any edge $(u, v) \in E_N$, the unlink prediction task on dynamic network aims to predict whether edge $(u, v)$ will disappear in $G_{N+1}$.
\end{definition}

\begin{figure}[!t]
	\centering
	\includegraphics[width=3.5in]{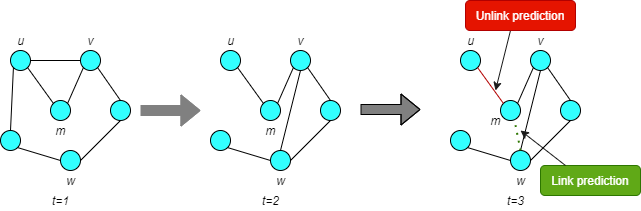}
	\caption{An example of temporal link prediction and unlink prediction. In practice, we use the link states at previous timestamps $t=1$ and $t=2$ to predict the formations and disappearances of links at timestamp $t=3$.}
	\label{fig_1}
\end{figure}

For example, as shown in Figure \ref{fig_1}, given a graph and its two snapshots at $t=1$ and $t=2$, we know all the changes of edges in both snapshots. For instance, nodes $v$ and $w$ are unconnected at timestamp $t=1$ and then get linked at timestamp $t=2$, while the edge between $u$ and $v$ presented at timestamp $t=1$ disappears at timestamp $t=2$. Then, we aim to predict all the to-be formed links (e.g., the edge between $m$ and $w$) and to-be removed links (e.g., the edge between $u$ and $m$) at timestamp $t=3$ by using the link states at previous timestamps.

\subsection{Random Walk Method}
\label{sec_ran}
Random walk is one of the most popular methods to measure the importance of nodes in the graph w.r.t. the given node. Suppose that we start a standard random walk from the given node $u$, $p_0$ is a one-hot vector in which the corresponding entry of $u$ is 1 and the rest entries are 0. $\textbf{N}_{rw}$ is the transition probability given as $\textbf{N}_{rw}=\textbf{D}^{-1}\textbf{A}$. Then, the $k$-steps probability vector can be iteratively calculated by
\begin{displaymath}
	p^{(k)} = \textbf{N}_{rw}^T p^{(k-1)} = (\textbf{N}_{rw}^T)^{k}p^{(0)}.
\end{displaymath}

Let $\textbf{P}^{(k)} \in \mathbb{R}^{|V| \times |V|}$ be the matrix that combines $p^{(k)}$ for all the nodes in the graph. Then, the $ij$-th entry in $\textbf{P}^{(k)}$ represents the probability that the random walker starts at node $i$ to reach node $j$ in $k$ steps. In this work, we use $\hat{\textbf{P}}^{(k)} = \textbf{P}^{(k)} + (\textbf{P}^{(k)})^T$ as the similarity scores for all the pairs of nodes, where $\hat{\textbf{P}}^{(k)}_{ij}$ denotes the sum of the probabilities that a random walker jumps from node $i$ to node $j$ as well as from node $j$ to node $i$ within $k$ steps.

\section{Methodology}
In this section, we illustrate the detailed procedures of LULS that contains three main steps. First, for each snapshot $G_t$, LULS calculates two matrices $\textbf{H}_t$ and $\textbf{W}_t$ by using local random walk methods to collect higher-order topological information. Then, LULS initializes a global representation matrix $\textbf{U}$ and a temporal representation matrix $\textbf{V}_t$ for each snapshot $G_t$ to represent the long-term relations and short-term relations respectively, and optimizes $\textbf{U}$ and $\textbf{V}_t$ by applying the method of NMF according to $\textbf{H}_t$ and $\textbf{W}_t$ of each snapshot. Finally, LULS computes the similarity matrix $\textbf{R}$ of the future snapshot based on the matrices $\textbf{V}_1, \textbf{V}_2, \cdots, \textbf{V}_N$ and $\textbf{U}$, and predicts the states of links and unlinks for the future snapshot. The whole framework is given in Algorithm \ref{alg_lupt}, and we elaborate each step as follows.

\begin{algorithm}[!h]
	\caption{The LULS Algorithm}
	\begin{algorithmic}[1]
		\Require A series of adjacency matrices $\textbf{A}_1, \textbf{A}_2, \cdots, \textbf{A}_N$, the hyper-parameters $\lambda$, $\gamma$, and $\theta$
		\Ensure Similarity matrix $\textbf{R}$ 
		\State compute matrices ${\textbf{H}_1}, {\textbf{H}_2}, \cdots, {\textbf{H}_N}$
		\State compute matrices ${\textbf{W}_1}, {\textbf{W}_2}, \cdots, {\textbf{W}_N}$
		\For{$t=1$ to $N$}
		\State randomly initialize ${\textbf{V}_t}$
		\EndFor
		\State randomly initialize $\textbf{U}$
		\Repeat:
		\For {$t=1$ to $N$}
		\State update ${\textbf{V}_t}$ according to Eq. (\ref{eqn:4})
		\EndFor
		\State update $\textbf{U}$ according to Eq. (\ref{eqn:5})
		\Until {termination criteria are reached}
		\State compute $\textbf{R}$ according to Eq. (\ref{eqn:3})
		\State \Return $\textbf{R}$
	\end{algorithmic}
	\label{alg_lupt}
\end{algorithm}

\subsection{Collecting Topological Information}
In the first step, for each snapshot $G_t$, LULS computes two matrices ${\textbf{H}_t}$ and ${\textbf{W}_t}$ to represent topological information. Note that the basic idea of NMF is to learn a low-dimensional vector for each node by factorizing the adjacency matrix. However, the adjacency matrix can not fully capture the higher-order relations between nodes. Moreover, since the real-world networks are always extremely large with millions or even billions of nodes, it is computationally expensive to apply global proximity methods to measure the strength between nodes. Therefore, a few variants of random walk, which can effectively collect the topological information, are proposed as solutions~\cite{Krylov_sub_app, kamuhanda2020sparse}.

In this work, we adopt light lazy random walk (LLRW) \cite{Krylov_sub_app} and propose a new random walk variant called modified light lazy random walk (MLLRW) to collect the higher-order topological information. LLRW and MLLRW are defined as follows:
\begin{itemize}
	\item [(1)] Light Lazy Random Walk (LLRW).
	\begin{displaymath}
		\textbf{N}_{rw}=(\textbf{D}+\alpha \textbf{I})^{-1}(\alpha \textbf{I}+\textbf{A}),
	\end{displaymath}
	where $\alpha \in N^{0+}$ is a hyper-parameter. Compared to standard random walk, LLRW retains some probability at the current node for the random walks. Moreover, LLRW degenerates to standard random walk when $\alpha =0$, and $\alpha = l$ indicates that the random walker performs $l$ loops on each node.
	
	\item [(2)] Modified Light Lazy Random Walk (MLLRW).
	\begin{displaymath}
		\textbf{N}_{rw}= \beta \textbf{S} + (1-\beta)\left((\textbf{D}+\alpha \textbf{I})^{-1}(\alpha \textbf{I}+\textbf{A})\right),
	\end{displaymath}
	where $\textbf{S}$ is the diagonal matrix with degree centrality similarity between every node in the network, and $\beta \in [0,1]$ is a hyper-parameter. MLLRW expects the ranking score of the nodes to be biased more to higher degree nodes, such that a random walker of MLLRW performs light lazy walk to one of the neighbors of the current node with probability $(1-\beta)$ and jumps to any other node in the graph according to $\textbf{S}$ with probability $\beta$. Additionally, MLLRW degenerates to LLRW when $\beta=0$.
\end{itemize}

Let $\hat{\textbf{P}}^{(k)}_{LLRW}$ and $\hat{\textbf{P}}^{(k)}_{MLLRW}$ represent the similarity matrices as introduced in Section \ref{sec_ran} obtained by performing LLRW and MLLRW over the graph, respectively. Then, we separately calculate $\textbf{H}_t = \hat{\textbf{P}}^{(k)}_{LLRW}$ and $\textbf{W}_t = \hat{\textbf{P}}^{(k)}_{MLLRW}$ for each snapshot $G_t$.

\subsection{Node Representation}
In the dynamic real-world network, nodes have long-term relations and short-term relations, and the temporal predictions of links and unlinks are determined by these two relations. Consequently, we leverage NMF to find the temporary representation matrices $\textbf{V}_1, \textbf{V}_2, \cdots, \textbf{V}_N$ and a global representation matrix $\textbf{U}$ such that the inner product of $\textbf{V}_i$ and $\textbf{U}$ can approximate the topological information $\textbf{W}_t$ of $G_t$. Note that, the global representation \textbf{U} can be considered as long-term relation between nodes, while the temporary representations $\textbf{V}_1, \textbf{V}_2, \cdots, \textbf{V}_N$ evolved over time is regarded as short-term relations.

Specifically, let $m$ be the dimension of node representation, and for each $\textbf{W}_t$, we aim to factorize the matrix as ${\textbf{W}_t} \approx \textbf{U} \textbf{V}_t^T$, where the matrices $\textbf{U} \in {\mathbb{R}_+}^{|V| \times m}$ and ${\textbf{V}_t} \in {\mathbb{R}_+}^{|V| \times m}$ represent the global representation matrix and temporary representation matrix, respectively. Additionally, to ensure that the temporary representations of nodes between snapshots do not change significantly, we apply regularization to maintain the smoothness. Therefore, for the matrix $\textbf{W}_t$, the objective function is given as
\begin{equation}\label{eqn:1}
	\begin{aligned}
		\mathop {\min }\limits_{\textbf{U} \ge 0,{\textbf{V}_t} \ge 0} J &= \sum\limits_{t = 1}^N {{\theta ^{N - t}}\left\| {{\textbf{W}_t} - \textbf{U} \textbf{V}_t^T} \right\|} _F^2 \\
		&+ \lambda \sum\limits_{t = 2}^N {\theta ^{N - t}}{{\left\| {{\textbf{V}_t} - {\textbf{V}_{t - 1}}} \right\|}^2_F} ,
	\end{aligned}
\end{equation}
where $\theta \in [0,1]$ is the decay weight and $\lambda$ is the smoothness weight.

Furthermore, to enhance the node representation $\textbf{U}$ that preserves more topological information,motivated by graph regularization technique~\cite{cai2010graph}, we define the similarity constraint term $M_{t}$ based on $\textbf{H}_t$ for each snapshot $G_t$ as follows
\begin{equation}
	M_{t} = \frac{1}{2}{\sum\limits_{ij} {\left\| {u_i - u_j} \right\|}^2} \textbf{H}_{t_{ij}} = Tr(\textbf{U}^T{\textbf{L}_t}\textbf{U}),
\end{equation}
where $\textbf{H}_{t_{ij}}$ is the $ij$-th entry in $\textbf{H}_t$, $\textbf{L}_t = \textbf{D}_t - \textbf{H}_t$ is the Laplacian matrix, and $Tr(\cdot)$ is the trace matrix. Overall, the enhanced objective function is expressed as follows:
\begin{equation}
	\label{eqn:2}
	\begin{aligned}
		\mathop {\min }\limits_{\textbf{U} \ge 0,\textbf{V}_t \ge 0} J =& \sum\limits_{t = 1}^N {{\theta ^{N - t}}\left\| {{\textbf{W}_t} - \textbf{U} \textbf{V}_t^T} \right\|} _F^2 + \gamma \sum\limits_{t = 1}^N M_t\\
		&+ \lambda \sum\limits_{t = 2}^N {\theta ^{N - t}}{{\left\| {{\textbf{V}_t} - {\textbf{V}_{t - 1}}} \right\|} ^2_F},
	\end{aligned}
\end{equation}
where $\gamma$ is a hyper-parameter to control the relevance importance of constraint terms.

\subsection{Optimization}
In this section, we present an iterative update algorithm to solve the optimization problem in Eq. (\ref{eqn:2}). In each iteration, the algorithm updates each matrix in turn while fixing the other matrices. This procedure repeats until the matrices converge or the maximum number of iterations is reached. By removing the irrelevant items, i.e., $\theta$, Eq. (\ref{eqn:2}) can be simplified as:
\begin{displaymath}
	\begin{aligned}
		\mathop {\min }\limits_{\textbf{U} \ge 0,{\textbf{V}_t} \ge 0} J &= {\sum\limits_{t = 1}^N {\left\| {{\textbf{W}_t} - \textbf{U} \textbf{V}_t^T} \right\|} ^2_F} + \gamma \sum\limits_{t = 1}^N {Tr({\textbf{U}^T}{\textbf{L}_t}\textbf{U})} \\
		& + \lambda \sum\limits_{t = 2}^N {{{\left\| {{\textbf{V}_t} - {\textbf{V}_{t - 1}}} \right\|}^2_F}}.
	\end{aligned}
\end{displaymath}

We first address the problem of optimizing $\textbf{V}_t$ for each $t \in [1,N]$. According to Eq. (\ref{eqn:2}), the optimization problem is transformed as follows:
\begin{equation}
	\begin{split}
		\mathop {\min}\limits_{{\textbf{V}_t} \ge 0} J =& \sum\limits_{t = 1}^N {Tr(\gamma ({\textbf{W}_t}}  - \textbf{U}\textbf{V}_t^T){({\textbf{W}_t} - \textbf{U}\textbf{V}_t^T)^T})
		\\& + \sum\limits_{t = 2}^{N - 1} {Tr(\lambda ({\textbf{V}_t}}  - {\textbf{V}_{t - 1}}){({\textbf{V}_t} - {\textbf{V}_{t - 1}})^T}).
	\end{split}
\end{equation}

Based on the non-negativity constraint of ${\textbf{V}_t}$ following the standard constraint optimization theory, we introduce the Lagrangian multiplier ${\phi _t} = \left[ {{\phi _{ij}}} \right]$ and minimize the Lagrangian function $L$, such that
\begin{displaymath}
	L = J + \sum\limits_{t = 1}^N {Tr({\phi _t}} {\textbf{V}_t}^T).
\end{displaymath}

By computing the derivate of $L$ with respect to ${\textbf{V}_t}$, we have the following expression:
\begin{displaymath}
	\begin{aligned}
		\frac{{\delta L}}{{\delta {\textbf{V}_t}}} &=  - 2{\textbf{W}_t}\textbf{U} + \textbf{U}\textbf{V}_t^T{\textbf{U}^T} + 2\lambda ({\textbf{V}_t} - {\textbf{V}_{t - 1}})+ {\phi _t}.
	\end{aligned}
\end{displaymath}
Next, by setting ${{\delta L} \over {\delta {\textbf{V}_t}}} = 0$ and ${\left[ {{\phi _t}} \right]_{ij}}{\left[ {{\textbf{V}_t}} \right]_{ij}} = 0$, we obtain the following equation for ${\left[ {{\textbf{V}_t}} \right]_{ij}}$:
\begin{displaymath}
	{[ - {\textbf{W}_t}\textbf{U} + \textbf{U}\textbf{V}_t^T{\textbf{U}^T} + \lambda ({\textbf{V}_t} - {\textbf{V}_{t - 1}}) ]_{ij}}{({\textbf{V}_t})_{ij}} = 0.
\end{displaymath}
Then, the update rule for ${\textbf{V}_t}$ is derived as follows:
\begin{equation}\label{eqn:4}
	{\textbf{V}_t} \leftarrow {\textbf{V}_t} \odot \sqrt {{{{\textbf{H}_t}\textbf{U} + \lambda ({\textbf{V}_{t - 1}})} \over {\textbf{U}\textbf{V}_t^T{\textbf{U}^T} + \lambda ({\textbf{V}_t})}}}.
\end{equation}

The global matrix $\textbf{U}$ can be learned in a very similar to latent factor ${\textbf{V}_t}$. To handle non-negative constraints, we introduce the Lagrangian multiplier
$\psi  = \left[ {{\psi _{ij}}} \right]$ and minimize the Lagrangian function $L$:
\begin{displaymath}
	L = J + \psi 
\end{displaymath}

Taking the derivate of L with respect to $U$, we have the following
expression
\begin{displaymath}
	\begin{aligned}
		{{\delta L} \over {\delta U}} =  - 2{W_t}{V_t} + 2U{V_t}^T{V_t}^T + 2\gamma {L_t}U + \psi ,
	\end{aligned}
\end{displaymath}
by setting ${{\delta L} \over {\delta {U}}} = 0$, and using KKT conditions 
${[\psi ]_{ij}}{[U]_{ij}} = 0$, we obtain the following equation for ${[U]_{ij}}$:
\[{[ - {W_t}{V_t} + U{V_t}^T{V_t}^T + \gamma {L_t}U]_{ij}}{({U})_{ij}} = 0\]

Then the update rule for $U$ is derived as follows:
\begin{equation}\label{eqn:5}
	U \leftarrow U \odot \sqrt {{{\sum\limits_{t = 1}^N {({W_t}{V_t} + \gamma {H_t}U)} } \over {\sum\limits_{t = 1}^N {(U{V_t}^T{V_t}^T + \gamma {D_t}U)} }}} .
\end{equation}

\subsection{Predicting Links and Unlinks}
After optimizing $\textbf{V}_1, \textbf{V}_2, \cdots, \textbf{V}_N$ and $\textbf{U}$, LULS calculates the similarity matrix of future snapshot for the predictions of links and unlinks. The probability of link formation and link disappearance can be obtained by the inner product of the latent factors $\textbf{U}$ and ${\textbf{V}_t}$ for each $t \in [1, N]$, such that 
\begin{equation}\label{eqn:3}
	\textbf{R} = \sum\nolimits_{t = 1}^N {\textbf{U}\textbf{V}_t}^T.
\end{equation}
Each $ij$-th entry in \textbf{R} denotes the proximity score for the pair of nodes $i$ and $j$. For the link prediction problem, the higher the similarity score between unconnected nodes in \textbf{R} the higher the probability of connect in the future. In contrast, for the unlink prediction problem, for two connected nodes in \textbf{R} if this pair of nodes have a lower similarity score, then they have a higher probability of disappearing in the future.

\subsection{Complexity Analysis}
\label{sec_com}
In this section, we discuss the time complexity of LULS. Local spectal diffusion \cite{a_sur_of_link} depends on the number of nodes on the network, and the time complexity is
$O(Nl{\left| V \right|^2})$, where $l$ is the number of iterations to convergence. For NMF, the computation is dominated by matrix multiplications, i.e., the matrix multiplication between $|V| \times |V|$ matrix and $|V| \times m$ matrix. Therefore, the complexity involved for updating $V_t$ and $U$ is $O(rN{\left| V \right|^2}m)$, where $r$ denotes the number of iterations. Besides, the complexity of computing $R$ is $O({\left| V \right|^2}m)$. Thus, the overall time complexity of our LULS model is
$O((rN + 1){\left| V \right|^2}m + Nl{\left| V \right|^2})$. Note that, since all matrices are sparse, the complexity between two sparse matrix is much smaller than $O({|V|^2})$.

\section{Experimental Setup}
\subsection{Datasets}
\label{sub_data}
We use six real-world dynamic networks for evaluating the performance of LULS. The datasets contain Facebook Forum, Reality Mining, Dublin, Hep-Th, Facebook Messages, and Retweet. Table \ref{tab_1} summarizes the detailed information of these datasets.

\begin{table*}[!t]
	\caption{Statistics of the datasets.}
	\label{tab_1}
	\centering
	\begin{tabular}{|l|c|c|c|c|c|c|}
		\hline
		~~  & Facebook Forum & Reality Mining & Dublin & Hep-Th & Facebook Messages & Haggle \\
		\hline
		Number of nodes & 899 & 6416 & 6454 & 22908 &274 & 18470\\
		\hline
		Number of edges & 7046 & 7250 & 24097 & 2444798 & 15737 & 2124 \\
		\hline
		Average degree & 15.68 & 2.26 & 7.45 & 213.44 & 16.57 &15.51 \\
		\hline
		Density & 0.0175 & 0.0004 & 0.0012 & 0.0093 & 0.0087 & 0.0568\\
		\hline
		Avg shortest-path distance & 2.8320 & 4.2367 & 6.6808 & 2.7220 & 3.0552 & 2.42\\
		\hline
		Number of snapshots & 6 & 7 & 8 & 11 & 8 & 6\\
		\hline
	\end{tabular}
\end{table*}
\begin{itemize}
	\item[-] \textbf{Facebook Forum}~\cite{nod_sca_fea}: The Facebook Forum network is the private messages exchanged between Facebook users from May to October in 2004, where the nodes are the users, and each edge is a message exchanged between a pair of users.
	\item[-] \textbf{Reality Mining}~\cite{nod_sca_fea}: Reality Mining network consists of individual's mobile phone call events between a set of core users at Massachusetts Institute of Technology (MIT), where the vertices are users, and each edge is a phone call or voicemail between a pair of users.
	\item[-] \textbf{Dublin}~\cite{nod_sca_fea}: Dublin network is a human contact network where the vertices represent individuals and the edges denote proximity. 
	\item[-] \textbf{Hep-Th\footnote{{http://networkrepository.com}}}: This dataset is a collaboration network from high energy physics theory section on arXiv, where vertices are authors and an edge denotes a common publication for a pair of authors.
	\item[-] \textbf{Facebook Messages\footnotemark[\value{footnote}]}: This dataset is a Facebook-like social network originated from online community of the students at the University of California, where the nodes are users, and the edges are the messages exchanged between pairs of users.
	\item[-] \textbf{Haggle\footnote{{http://konect.cc/networks}}}: This  network reflects connections between people as measured by wireless devices carried by the participants. A node symbolizes a person, and an edge between two people indicates that they came into touch.
\end{itemize}

\subsection{Evaluation Metric}
\label{sec_data}
In this paper, we apply AUC and average precision (AP) to evaluate the performance of LULS. Specifically, for each dataset, we use ${G_1},{G_2}, \cdots, {G_{N - 1}}$ as the training data and ${G_N}$ as the test data. Furthermore, the test data is divided into positive test samples and negative test samples. For the link prediction problem, the positive test set consists of the edges that appear in ${G_N}$ and do not present in ${G_{N - 1}}$, while the negative test set consists of the edges that do not appear in ${G_{N - 1}}$ and ${G_N}$. On the other hand, for unlink prediction problem, the positive test set contains the edges that appear in ${G_{N - 1}}$ and ${G_N}$, while the negative test set consists of the edges that present in ${G_{N - 1}}$ and disappear in ${G_N}$. To avoid the class imbalance, we randomly generate the same size of negative test set as that of positive test set for both link prediction and unlink prediction. In addition, the experiments are carried out five times independently and the average result is reported.

Generally, the AUC is described as the likelihood that the randomly selected actual link in the positive test set is
assigned a higher score than a randomly selected link in the negative test set.
Formally, if among $n$ comparisons, there are ${n'}$ times the the edges in the negative test set  has a lower score than the edges in the positive test set and
$n''$ times they have the same scores, the AUC scores are calculated as follows 
\begin{displaymath}
	\begin{aligned}
		AUC = \frac{{n' + 0.5*n''}}{n}.
	\end{aligned}
\end{displaymath}
Note that, an algorithm has a better performance than pure chance when the value of AUC is bigger than 0.5.

AP combines recall and precision for ranking results. We calculate the precision after each true positive given a ranked list of predicted links. The average of these values gives the average precision for that link.

\subsection{Baselines}
We compare our method with the the-state-of-the-art methods as follows:
\begin{itemize}
	\item[-] \textbf{AA}~\cite{pre_mis_lin}: AA assumes that two nodes are more likely to be linked together if they share more common neighbors.
	\item[-] \textbf{DCN}~\cite{tian2020exploiting}: This method uses a decay common neighbor to characterize the relationship between node pair. 
	\item[-] \textbf{TD}~\cite{dunlavy2011temporal}: This method stacks all adjacency matrices of historical snapshots into a tensor with the time as the third dimension to improve the link prediction results.
	\item[-] \textbf{TMF}~\cite{yu2017temporally}: This method uses matrix factorization techniques to characterize the network characteristics as a function of time
	\item[-] \textbf{GrNMF}~\cite{ma2018graph}: This method directly approximates the link matrix over time $T$ using NMF by setting networks from 1 to $T-1$ as a regularizer.
\end{itemize}

Note that, the methods AA and DCN can be applied only to static network. Therefore, in the case of link prediction, these approaches are performed based on the links from all past time periods by combining them into a single link matrix. For the unlink prediction task, these approaches are conducted based on the link states over ${G_{N - 1}}$.

\section{Experimental Results}
In the experiments, all the parameters of LULS have been manually tuned. Specifically, we set $m = 5$, $\theta  = 0.4$, $\gamma  = 1$ and $\lambda = 0.0001$ for facebook Forum, Facebook Messages and Haggle networks. And for Reality mining and Dublin network, we set $\gamma  = 0.0001$. For the random walks variants, we use $\alpha = 1$ for LLRW, and $\beta  = 0.01$ for MLLRW. Besides, we use the random walk step $k = 4$ for the Facebook Forum, Reality mining, Facebook Messages and Hep-Th networks, and $k = 5$ for the Dublin network. Moreover, for evaluating the effectiveness of smoothness and constraint terms, we implement three versions of LULS as follows:
\begin{itemize}
	\item[-] LULS$_1$: $\lambda \neq 0$ and $\gamma \neq 0$;
	\item[-] LULS$_2$: $\lambda \neq 0$ and $\gamma = 0$;
	\item[-] LULS$_3$: $\lambda = 0$ and $\gamma = 0$.
\end{itemize}

\begin{table*}[!t]
	\caption{The AUC scores for the link prediction task.}
	\label{tab_2}
	\centering
	\scalebox{0.9}{
		\begin{tabular}{|c|c|c|c|c|c|c|}
			\hline
			\diagbox{Methods}{Datasets} & Facebook Forum & Reality Mining & Dublin & Hep-Th & Facebook Messages & Haggle \\
			\hline
			LULS$_{1}$ & \textbf{0.9324} & 0.9735 & \textbf{0.9913} & \textbf{0.7351} & \textbf{0.9765} & \textbf{0.9881} \\
			\hline
			LULS$_{2}$ & 0.9232 & \textbf{0.9764} & 0.9909 & 0.7089 & 0.9741 & 0.9870  \\
			\hline
			LULS$_{3}$ & 0.8278 & 0.9750 & 0.9908 & 0.6974 & 0.9750 & 0.9835\\
			\hline
			AA & 0.5252 & 0.5192 & 0.8914 & 0.5612 & 0.7351 & 0.9230\\
			\hline
			DCN & 0.5313 & 0.5194 & 0.9620 & 0.5612 & 0.4601 & 0.4532 \\
			\hline
			TD & 0.9059 & 0.9214 & 0.6495 & 0.6268 & 0.9192 & 0.5693 \\
			\hline
			TMF & 0.8314 & 0.9284 & 0.6037 & 0.6985 & 0.7122 & 0.9127 \\
			\hline
			GrNMF & 0.8792 & 0.9204 & 0.6542 & 0.6837 & 0.9474 & 0.9213  \\
			\hline
	\end{tabular}}
\end{table*}

\begin{table*}[!t]
	\caption{The AP scores for the Link prediction task.}
	\label{tab_4}
	\centering
	\scalebox{0.9}{
		\begin{tabular}{|c|c|c|c|c|c|c|}
			\hline
			\diagbox{Methods}{Datasets} & Facebook Forum & Reality Mining & Dublin & Hep-Th & Facebook Messages & Haggle \\
			\hline
			LULS$_{1}$ &\textbf{ 0.8942} &0.9570 & \textbf{0.9850}& \textbf{0.7012} & \textbf{0.9528} &\textbf{0.9810} \\
			\hline
			LULS$_{2}$ & 0.8731 & \textbf{0.9604} & 0.9843 & 0.6926 & 0.9521& 0.9775\\
			\hline
			LULS$_{3}$ & 0.8646 & 0.9602 & 0.9844  & 0.6845 & 0.9510 & 0.9642\\
			\hline
			AA & 0.5160 & 0.5150 & 0.8909 & 0.5547 & 0.7400 & 0.9430 \\
			\hline
			DCN & 0.5646 & 0.4424 & 0.9563 & 0.5610 & 0.4619 & 0.6096 \\
			\hline
			TD & 0.8870 & 0.9106 & 0.9028 & 0.6248 & 0.9357 & 0.7481 \\
			\hline
			TMF & 0.8271 & 0.9117 & 0.5926 & 0.6954 & 0.7018 & 0.8020 \\
			\hline
			GrNMF & 0.8444 & 0.9071 & 0.6018 & 0.6773 & 0.7155 & 0.8587 \\
			\hline
	\end{tabular}}
\end{table*}

\begin{table*}[!t]
	\caption{The AUC scores for the unlink prediction task.}
	\label{tab_3}
	\centering
	\scalebox{0.9}{
		\begin{tabular}{|c|c|c|c|c|c|c|}
			\hline
			\diagbox{Methods}{Datasets} & Facebook Forum & Reality Mining & Dublin & Hep-Th & Facebook Messages & Haggle \\
			\hline
			LULS$_{1}$ &\textbf{ 0.8363} &\textbf {0.7970} & 0.7525& \textbf{0.6923} & 0.7790 &0.9345 \\
			\hline
			LULS$_{2}$ & 0.8268 & 0.7968 & \textbf{0.7532} & 0.6567 & \textbf{0.8044}& 0.9334\\
			\hline
			LULS$_{3}$ & 0.8278 & 0.7661 & 0.7517  & 0.5051 & 0.8036 & 0.9316\\
			\hline
			AA & 0.546 & 0.5192 & 0.5437 & 0.5558 & 0.5101 & 0.6492 \\
			\hline
			DCN & 0.4858 & 0.5014 & 0.5871 & 0.5437 & 0.5117 & 0.6096 \\
			\hline
			TD & 0.7596 & 0.7651 & 0.5731 & 0.6154 & 0.7032 & \textbf{0.9367} \\
			\hline
			TMF & 0.6780 & 0.7946 & 0.5861 & 0.6748 & 0.6961 & 0.8143 \\
			\hline
			GrNMF & 0.6395 & 0.7848 & 0.6351 & 0.6645 & 0.7411 & 0.8445 \\
			\hline
	\end{tabular}}
\end{table*}

\begin{table*}[!t]
	\caption{The AP scores for the Unlink prediction task.}
	\label{tab_5}
	\centering
	\scalebox{0.9}{
		\begin{tabular}{|c|c|c|c|c|c|c|}
			\hline
			\diagbox{Methods}{Datasets} & Facebook Forum & Reality Mining & Dublin & Hep-Th & Facebook Messages & Haggle \\
			\hline
			LULS$_{1}$ &\textbf{ 0.8120} &0.7885 & \textbf{0.8231}& \textbf{0.7014} & 0.8162 &0.9333 \\
			\hline
			LULS$_{2}$ & 0.7958 & 0.8773 & 0.8218 & 0.6612 & \textbf{0.8293}& 0.9320\\
			\hline
			LULS$_{3}$ & 0.7877 & \textbf {0.8776} & 0.8210  & 0.5152 & 0.8280 & 0.9268\\
			\hline
			AA & 0.5462 & 0.5083 & 0.5437 & 0.5610 & 0.5089 & 0.6520 \\
			\hline
			DCN & 0.4931 & 0.5011 & 0.5667 & 0.4619 & 0.4991 & 0.5741 \\
			\hline
			TD & 0.8077 & 0.7620 & 0.7447 & 0.6272 & 0.7928 & \textbf{0.9377} \\
			\hline
			TMF & 0.8280 & 0.7940 & 0.5612 & 0.6675 & 0.6961 & 0.8138 \\
			\hline
			GrNMF & 0.7812 & 0.7845 & 0.6326 & 0.6651 & 0.7155 & 0.8341 \\
			\hline
	\end{tabular}}
\end{table*}

\subsection{Link Prediction}
In this experiment, we evaluate the performance of LULS for link prediction. Table \ref{tab_2} and Table \ref{tab_4} shows the performance of different approaches on six dynamic networks for the link prediction task. It can be observed that LULS models perform best in all the datasets, indicating that LULS can effectively integrate both temporal and structural information to extract significant node representation for the link prediction task. Precisely, our model has shown an impressive performance even on a very sparse graphs, e.g., Dublin network. Furthermore, all the approaches based on dynamic characteristics (i.e., LULS, TD, TMF, and GrNMF) consistently perform better than the approaches that ignore the temporal behaviour of the network (i.e., AA and DCN) in almost all the datasets. In addition, among LULS models, LULS$_1$ is better than LULS$_2$ and LULS$_3$. Consequently, the smoothness and similarity constraint terms play important roles in link prediction.

\subsection{Unlink Prediction}
Next, we investigate the effectiveness of LULS for the unlink prediction task. In Table \ref{tab_3} and Table \ref{tab_5}, the AUC and AP scores show that LULS models significantly outperform the baseline methods in almost all the datasets. Precisely, our model have shown an impressive performance on Haggle network, suggesting that unlink prediction is more likely on dense graph than on sparse graphs. Similar to link prediction task, all the approaches which consider temporal information of the network outperform other methods. Moreover, among LULS models, LULS$_1$ is also better than LULS$_2$ and LULS$_3$, which indicates that the smoothness and similarity constraint terms are useful in unlink prediction. Additionally, comparing to link prediction, we can observe that the AUC scores of unlink prediction are lower, indicating that it is more difficult to predict unlinks than links in the future network.

\section{Conclusion}
In this work, we address the problems of temporal link prediction and unlink prediction on dynamic networks. Assuming that there are two kinds of relations between nodes, namely long-term relation and short-term relation, we propose an effective algorithm called LULS for temporal link prediction and unlink prediction based on such relations. Specifically, LULS collects the topological information for each snapshot of a dynamic network and generates a global matrix and a sequence of temporary matrices to represent the long-term relation and short-term relation. Then, LULS utilizes the global matrix and the temporary matrices to predict the links and unlinks for the future network. The experiments conducted on six real-world networks show the superior results of LULS compared with the state of the art methods.


%

\bibliographystyle{IEEEtran}
\bibliography{references.bib}

\begin{thebibliography}{10}
\providecommand{\url}[1]{#1}
\csname url@samestyle\endcsname
\providecommand{\newblock}{\relax}
\providecommand{\bibinfo}[2]{#2}
\providecommand{\BIBentrySTDinterwordspacing}{\spaceskip=0pt\relax}
\providecommand{\BIBentryALTinterwordstretchfactor}{4}
\providecommand{\BIBentryALTinterwordspacing}{\spaceskip=\fontdimen2\font plus
\BIBentryALTinterwordstretchfactor\fontdimen3\font minus
  \fontdimen4\font\relax}
\providecommand{\BIBforeignlanguage}[2]{{%
\expandafter\ifx\csname l@#1\endcsname\relax
\typeout{** WARNING: IEEEtran.bst: No hyphenation pattern has been}%
\typeout{** loaded for the language `#1'. Using the pattern for}%
\typeout{** the default language instead.}%
\else
\language=\csname l@#1\endcsname
\fi
#2}}
\providecommand{\BIBdecl}{\relax}
\BIBdecl

\bibitem{kumar2020link}
A.~Kumar, S.~S. Singh, K.~Singh, and B.~Biswas, ``Link prediction techniques,
  applications, and performance: A survey,'' \emph{Physica A: Statistical
  Mechanics and its Applications}, vol. 553, p. 124289, 2020.

\bibitem{mat_fac_tec}
Y.~{Koren}, R.~{Bell}, and C.~{Volinsky}, ``Matrix factorization techniques for
  recommender systems,'' \emph{IEEE Computer}, vol.~42, no.~8, pp. 30--37,
  2009.

\bibitem{a_rev_of_rel}
M.~{Nickel}, K.~{Murphy}, V.~{Tresp}, and E.~{Gabrilovich}, ``A review of
  relational machine learning for knowledge graphs,'' \emph{arXiv preprint
  arXiv:1503.00759}, vol. 104, no.~1, pp. 11--33, 2016.

\bibitem{net_bas_pre_of}
I.~A. {Kovács}, K.~{Luck}, K.~{Spirohn}, Y.~{Wang}, C.~{Pollis},
  S.~{Schlabach}, W.~{Bian}, D.~K. {Kim}, N.~{Kishore}, T.~{Hao}, M.~A.
  {Calderwood}, M.~{Vidal}, and A.~L. {Barabási}, ``Network-based prediction
  of protein interactions.'' \emph{Nature Communications}, vol.~10, no.~1,
  2019.

\bibitem{evo_net_ana}
C.~{Aggarwal} and K.~{Subbian}, ``Evolutionary network analysis: A survey,''
  \emph{ACM Computing Surveys}, vol.~47, no.~1, 2014.

\bibitem{a_sur_of_link}
V.~{Martínez}, F.~{Berzal}, and J.-C. {Cubero}, ``A survey of link prediction
  in complex networks,'' \emph{ACM Computing Surveys}, vol.~49, no.~4, 2017.

\bibitem{link_pre_in_soc}
P.~{Wang}, B.~{Xu}, Y.~{Wu}, and X.~{Zhou}, ``Link prediction in social
  networks: the state-of-the-art,'' \emph{Science in China Series F:
  Information Sciences}, vol.~58, no.~1, pp. 1--38, 2015.

\bibitem{the_link_pre}
D.~{Liben-Nowell} and J.~{Kleinberg}, ``The link-prediction problem for social
  networks,'' \emph{Journal of the Association for Information Science and
  Technology}, vol.~58, no.~7, pp. 1019--1031, 2007.

\bibitem{chen2018exploiting}
H.~Chen and J.~Li, ``Exploiting structural and temporal evolution in dynamic
  link prediction,'' in \emph{Proceedings of the 27th ACM International
  Conference on Information and Knowledge Management}, 2018, pp. 427--436.

\bibitem{deng2016latent}
D.~Deng, C.~Shahabi, U.~Demiryurek, L.~Zhu, R.~Yu, and Y.~Liu, ``Latent space
  model for road networks to predict time-varying traffic,'' in
  \emph{Proceedings of the 22nd ACM SIGKDD International Conference on
  Knowledge Discovery and Data Mining}, 2016, pp. 1525--1534.

\bibitem{dunlavy2011temporal}
D.~M. Dunlavy, T.~G. Kolda, and E.~Acar, ``Temporal link prediction using
  matrix and tensor factorizations,'' \emph{ACM Transactions on Knowledge
  Discovery from Data (TKDD)}, vol.~5, no.~2, pp. 1--27, 2011.

\bibitem{rossi2013modeling}
R.~A. Rossi, B.~Gallagher, J.~Neville, and K.~Henderson, ``Modeling dynamic
  behavior in large evolving graphs,'' in \emph{Proceedings of the sixth ACM
  international conference on Web search and data mining}, 2013, pp. 667--676.

\bibitem{sarkar2005dynamic}
P.~Sarkar and A.~W. Moore, ``Dynamic social network analysis using latent space
  models,'' \emph{Acm Sigkdd Explorations Newsletter}, vol.~7, no.~2, pp.
  31--40, 2005.

\bibitem{yu2017temporally}
W.~Yu, C.~C. Aggarwal, and W.~Wang, ``Temporally factorized network modeling
  for evolutionary network analysis,'' in \emph{Proceedings of the Tenth ACM
  International Conference on Web Search and Data Mining}, 2017, pp. 455--464.

\bibitem{cai2008non}
D.~Cai, X.~He, X.~Wu, and J.~Han, ``Non-negative matrix factorization on
  manifold,'' in \emph{2008 Eighth IEEE International Conference on Data
  Mining}.\hskip 1em plus 0.5em minus 0.4em\relax IEEE, 2008, pp. 63--72.

\bibitem{fra_onl_rel}
H.~{Kwak}, H.~{Chun}, and S.~{Moon}, ``Fragile online relationship: a first
  look at unfollow dynamics in twitter,'' in \emph{Proceedings of the SIGCHI
  Conference on Human Factors in Computing Systems}, 2011, pp. 1091--1100.

\bibitem{loo_fri_on_fac}
D.~{Quercia}, M.~{Bodaghi}, and J.~{Crowcroft}, ``Loosing ''friends'' on
  facebook,'' in \emph{Proceedings of the 4th Annual ACM Web Science Conference
  on}, 2012.

\bibitem{the_imp_of_net}
F.~{Kivran-Swaine}, P.~{Govindan}, and M.~{Naaman}, ``The impact of network
  structure on breaking ties in online social networks: unfollowing on
  twitter,'' in \emph{Proceedings of the SIGCHI Conference on Human Factors in
  Computing Systems}, 2011, pp. 1101--1104.

\bibitem{pre_and_ide_mis}
R.~{Eyal}, A.~{Rosenfeld}, S.~{Sina}, and S.~{Kraus}, ``Predicting and
  identifying missing node information in social networks,'' \emph{ACM
  Transactions on Knowledge Discovery From Data}, vol.~8, no.~3, 2014.

\bibitem{pre_mis_lin}
T.~{Zhou}, L.~{Lü}, and Y.-C. {Zhang}, ``Predicting missing links via local
  information,'' \emph{European Physical Journal B}, vol.~71, no.~4, pp.
  623--630, 2009.

\bibitem{the_bur_dyn_of}
S.~A. {Myers} and J.~{Leskovec}, ``The bursty dynamics of the twitter
  information network,'' in \emph{Proceedings of the 23rd international
  conference on World wide web}, 2014, pp. 913--924.

\bibitem{sem_unl_pre}
M.~A. de~{Oliveira}, K.~C. {Revoredo}, and J.~E.~O. {Luna}, ``Semantic unlink
  prediction in evolving social networks through probabilistic description
  logic,'' in \emph{2014 Brazilian Conference on Intelligent Systems}, 2014.

\bibitem{Krylov_sub_app}
K.~{He}, P.~{Shi}, D.~{Bindel}, and J.~E. {Hopcroft}, ``Krylov subspace
  approximation for local community detection in large networks,'' \emph{ACM
  Transactions on Knowledge Discovery From Data}, vol.~13, no.~5, pp. 1--30,
  2019.

\bibitem{link_pre_in_dyn}
M.~{Rahman} and M.~A. {Hasan}, ``Link prediction in dynamic networks using
  graphlet,'' in \emph{ECML PKDD 2016 European Conference on Machine Learning
  and Knowledge Discovery in Databases - Volume 9851}, 2016, pp. 394--409.

\bibitem{link_pre_usi}
İsmail {Güneş}, Şule {Gündüz-Öğüdücü}, and Z.~{Çataltepe}, ``Link
  prediction using time series of neighborhood-based node similarity scores,''
  \emph{Data Mining and Knowledge Discovery}, vol.~30, no.~1, pp. 147--180,
  2016.

\bibitem{moradabadi2017novel}
B.~Moradabadi and M.~R. Meybodi, ``A novel time series link prediction method:
  Learning automata approach,'' \emph{Physica A: Statistical Mechanics and its
  Applications}, vol. 482, pp. 422--432, 2017.

\bibitem{acar2009link}
E.~Acar, D.~M. Dunlavy, and T.~G. Kolda, ``Link prediction on evolving data
  using matrix and tensor factorizations,'' in \emph{2009 IEEE International
  conference on data mining workshops}.\hskip 1em plus 0.5em minus 0.4em\relax
  IEEE, 2009, pp. 262--269.

\bibitem{zhu2016scalable}
L.~Zhu, D.~Guo, J.~Yin, G.~Ver~Steeg, and A.~Galstyan, ``Scalable temporal
  latent space inference for link prediction in dynamic social networks,''
  \emph{IEEE Transactions on Knowledge and Data Engineering}, vol.~28, no.~10,
  pp. 2765--2777, 2016.

\bibitem{chi2007evolutionary}
Y.~Chi, X.~Song, D.~Zhou, K.~Hino, and B.~L. Tseng, ``Evolutionary spectral
  clustering by incorporating temporal smoothness,'' in \emph{Proceedings of
  the 13th ACM SIGKDD international conference on Knowledge discovery and data
  mining}, 2007, pp. 153--162.

\bibitem{lin2008facetnet}
Y.-R. Lin, Y.~Chi, S.~Zhu, H.~Sundaram, and B.~L. Tseng, ``Facetnet: a
  framework for analyzing communities and their evolutions in dynamic
  networks,'' in \emph{Proceedings of the 17th international conference on
  World Wide Web}, 2008, pp. 685--694.

\bibitem{yu2017link}
W.~Yu, W.~Cheng, C.~C. Aggarwal, H.~Chen, and W.~Wang, ``Link prediction with
  spatial and temporal consistency in dynamic networks.'' in \emph{IJCAI},
  2017, pp. 3343--3349.

\bibitem{chen2019lstm}
J.~Chen, J.~Zhang, X.~Xu, C.~Fu, D.~Zhang, Q.~Zhang, and Q.~Xuan, ``E-lstm-d: A
  deep learning framework for dynamic network link prediction,'' \emph{IEEE
  Transactions on Systems, Man, and Cybernetics: Systems}, 2019.

\bibitem{yang2019advanced}
M.~Yang, J.~Liu, L.~Chen, Z.~Zhao, X.~Chen, and Y.~Shen, ``An advanced deep
  generative framework for temporal link prediction in dynamic networks,''
  \emph{IEEE transactions on cybernetics}, vol.~50, no.~12, pp. 4946--4957,
  2019.

\bibitem{lei2019gcn}
K.~Lei, M.~Qin, B.~Bai, G.~Zhang, and M.~Yang, ``Gcn-gan: A non-linear temporal
  link prediction model for weighted dynamic networks,'' in \emph{IEEE INFOCOM
  2019-IEEE Conference on Computer Communications}.\hskip 1em plus 0.5em minus
  0.4em\relax IEEE, 2019, pp. 388--396.

\bibitem{str_dyn_of_kno}
J.~{Preusse}, J.~{Kunegis}, M.~{Thimm}, S.~{Staab}, and T.~{Gottron},
  ``Structural dynamics of knowledge networks,'' in \emph{Seventh International
  AAAI Conference on Weblogs and Social Media}, 2013.

\bibitem{kamuhanda2020sparse}
D.~Kamuhanda, M.~Wang, and K.~He, ``Sparse nonnegative matrix factorization for
  multiple-local-community detection,'' \emph{IEEE Transactions on
  Computational Social Systems}, vol.~7, no.~5, pp. 1220--1233, 2020.

\bibitem{cai2010graph}
D.~Cai, X.~He, J.~Han, and T.~S. Huang, ``Graph regularized nonnegative matrix
  factorization for data representation,'' \emph{IEEE transactions on pattern
  analysis and machine intelligence}, vol.~33, no.~8, pp. 1548--1560, 2010.

\bibitem{nod_sca_fea}
A.~{Grover} and J.~{Leskovec}, ``node2vec: Scalable feature learning for
  networks,'' in \emph{Proceedings of the 22nd ACM SIGKDD International
  Conference on Knowledge Discovery and Data Mining}, vol. 2016, 2016, pp.
  855--864.

\bibitem{tian2020exploiting}
H.~Tian and R.~Zafarani, ``Exploiting common neighbor graph for link
  prediction,'' in \emph{Proceedings of the 29th ACM International Conference
  on Information \& Knowledge Management}, 2020, pp. 3333--3336.

\bibitem{ma2018graph}
X.~Ma, P.~Sun, and Y.~Wang, ``Graph regularized nonnegative matrix
  factorization for temporal link prediction in dynamic networks,''
  \emph{Physica A: Statistical mechanics and its applications}, vol. 496, pp.
  121--136, 2018.

\end{thebibliography}

\vspace{22pt}
\begin{IEEEbiography}[{\includegraphics[width=1in,height=1.25in,clip,keepaspectratio]{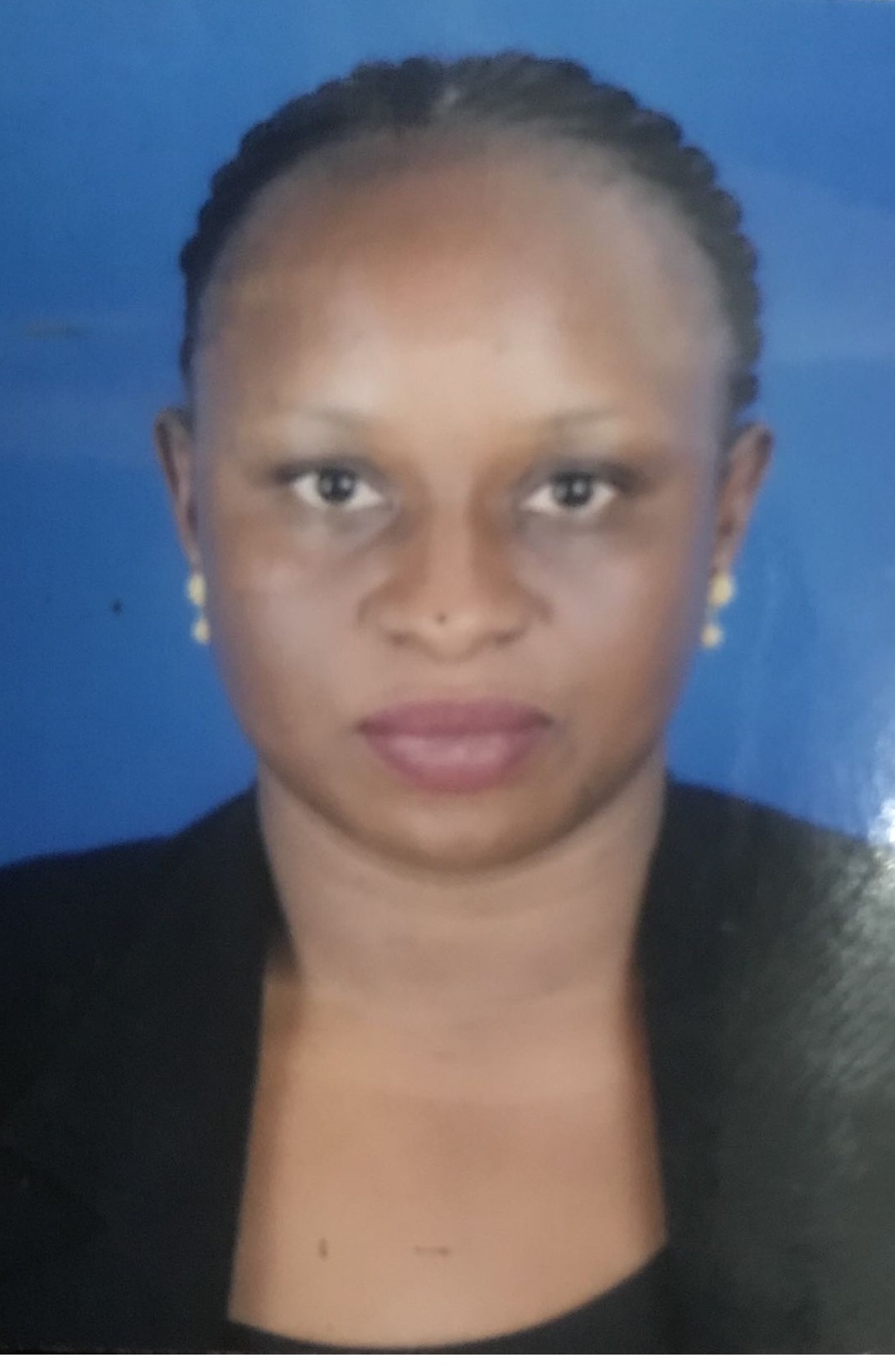}}]{Christina Muro} received  bachelor's degree with Computer Science from the University of Dar-es-salaam, Dar-es-salaam, Tanzania in 2009 and the master's degree in Computer Science, from the University of Dodoma, Dodoma, Tanzania in 2013. She is currently pursuing the Ph.D. degree with the School of
Computer Science and Technology, Huazhong University of Science and Technology, Wuhan, China.
She has been a Lecturer in the Department of Computer Science and Engineering at the University of Dodoma since 2018. Her research interests include machine learning and social network analysis. 
\end{IEEEbiography}
 
\vspace{22pt}
\begin{IEEEbiography}[{\includegraphics[width=1in,height=1.25in,clip,keepaspectratio]{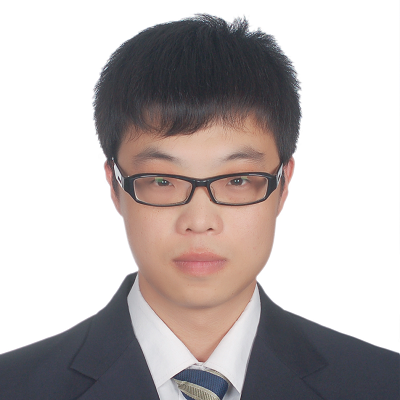}}]{Boyu Li}
received his M.S.\ degree and Ph.D.\ degree in Computer Science and Technology from Jilin University, in 2014 and 2018, respectively. He is currently a post-doctor in School of Computer Science and Technology, Huazhong University of Science and Technology. His research interests include privacy-preserving data publishing, social network and graph embedding.
\end{IEEEbiography}

\vspace{22pt}
\begin{IEEEbiography}[{\includegraphics[width=1in,height=1.25in,clip,keepaspectratio]{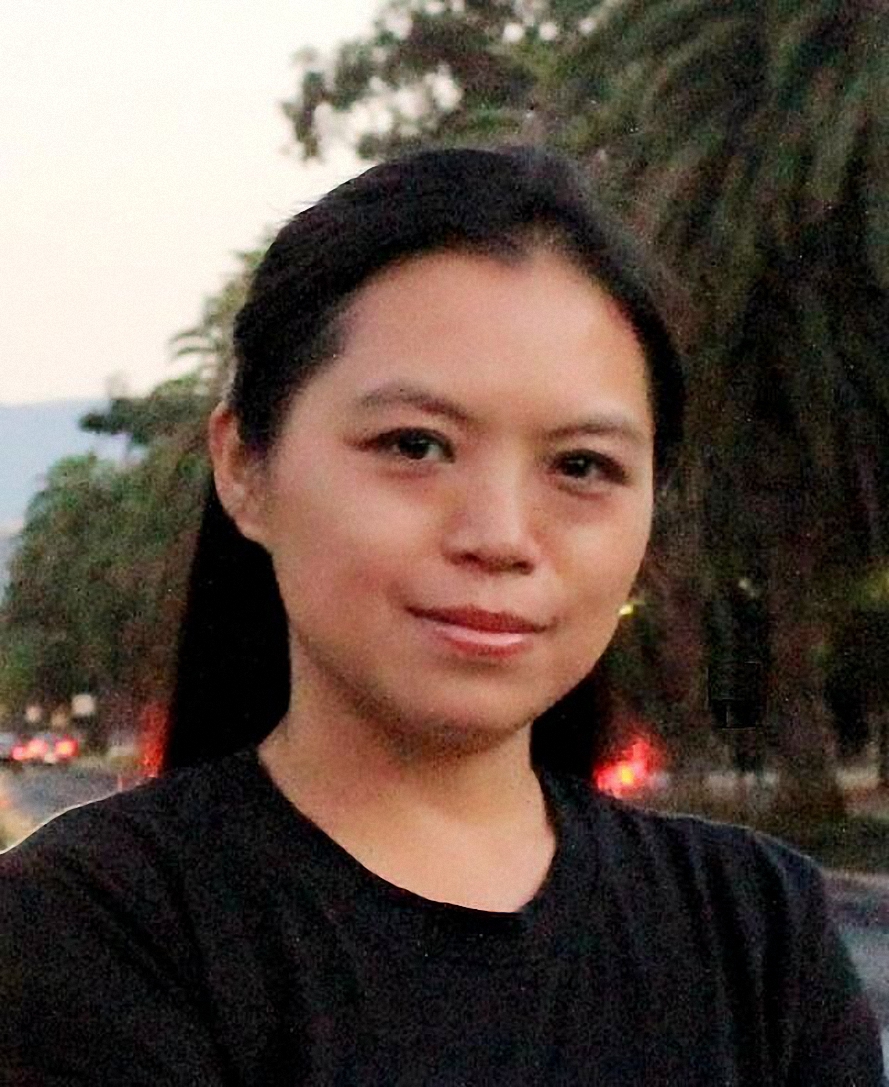}}]{Kun He}
is currently a Professor in School of Computer Science and Technology, Huazhong University of Science and Technology (HUST) Wuhan, P.R. China; and a Mary Shepard B. Upson Visiting Professor for the 2016-2017 Academic year in Engineering, Cornell University NY, USA. She received the B.S degree in physics from Wuhan University, Wuhan, China, in 1993; the M.S. degree in computer science from Huazhong Normal University, Wuhan, China, in 2002; and the Ph.D. degree in system engineering from HUST, Wuhan, China, in 2006. Her research interests include machine learning, deep learning, social networks, and algorithm design.
\end{IEEEbiography}


\vfill

\end{document}